\documentclass[12pt]{article}
\begin{document}
\def\p {{\partial}}
\def\n {{\nu}}
\def\m {{\mu}}
\def\a {{\alpha}}
\def\bt {{\beta}}
\def\f {{\phi}}
\def\th {{\theta}}
\def\g {{\gamma}}
\def\eps {{\epsilon}}
\def\e {{\psi}}
\def\k {{\chi}}
\def\la {{\lambda}}
\def\na {{\nabla}}
\def\bn {\begin{eqnarray}}
\def\en {\end{eqnarray}}
\centerline{{\bf {\large Hamilton-Jacobi quantization of singular
Lagrangians}}} \vspace{.3cm} \centerline{{\bf {\large with linear
velocities}}} \vspace {3cm} \centerline {\bf {SAMI I. MUSLIH }}
\vspace {0.3cm} \centerline {Department of Physics} \vspace
{0.1cm}\centerline {Al-Azhar University-Gaza, Palestine}
\vspace{2cm} \centerline {\bf {HOSAM A. El-ZALAN }} \vspace{0.3cm}
\centerline {Department of mathematics}\vspace {0.1cm} \centerline
{Al-Aqsa University-Gaza, Palestine}\vspace{2cm} \centerline {\bf
{EQAB M. RABEI }} \vspace{0.3cm} \centerline {Department of
Physics}\vspace {0.1cm} \centerline {Mu'tah University, Al-Karak,
Jordan} \vspace {2.5cm} \

 {{\bf ABSTRACT}}
\vspace{0.7cm}

In this paper, constrained Hamiltonian systems with linear
velocities are investigated by using the Hamilton-Jacobi method.
We shall consider the integrablity conditions on the equations of
motion and the action function as well in order to obtain the path
integral quantization of singular Lagrangians with linear
velocities.

\newpage
\section{Introduction} \vspace{0.3cm}

The study of singular Lagrangian with linear velocity has been
dealt within the last $50$ years by Dirac's Hamiltonian formulism
[1, 2]. In this formalism Dirac showed that the algebra of Poisson
brackets determines a division of constraints into two classes:
so-called first-class and second-class constraints. The
first-class constraints are those that have zero Poisson brackets
with all other constraints in the subspace of phase space in which
constraints hold; constraints which are not first-class are by
definition second-class. Also in his method, the Poisson brackets
in a second class constraints systems are converted into Dirac
brackets to attain self-consistency. However, wherever we adopt
the Dirac method, we frequently meet the problem of the operating
ordering ambiguity. Besides, the presence of first class
constraints in such theories requires care when applying Dirac's
method, since the first class constraints are the generators of
gauge transformations which lead to the gauge freedom. In other
words, the equations of motion are still degenerate and depend on
the functional arbitraries, one has to impose external gauge
fixing constraint for each first class constraint which is not
always an easy task.

\vspace{0.5cm}

Recently, G\"{u}ler [3-6] have proposed an alternating approach to
constrained systems that avoids the separations of constraints
into first and second class and the use of weak and strong
equations. This new method of analysis has been successfully used
by many authors [7-9] and is by now a standard technique to deal
with constrained system. Besides, the canonical path integral
method based on the Hamilton-Jacobi method have been initiated in
[10] to obtain that path integral quantization as an integration
over the canonical phase space coordinate without any need to use
any gauge fixing conditions [7] as well as, no need to enlarge the
initial phase-space by introducing unphysical auxiliary fields
[11,12].

\vspace{0.5cm}

Some authors [9, 13-16] have investigated singular Lagrangians
with linear velocities using Dirac's procedure. For example, using
the Lagrangian given in reference [15], there are different
approaches which give different results. Besides, in Ref. [13] the
authors have investigated singular Lagrangian with linear
velocities without considering the integrability condition on the
action function. On the other hand in reference [9], the authors
have investigated singular Lagrangian with linear velocities by
using the Hamilton-Jacobi method and obtained the integrable
action directly without considering the total variation of
constraints.

\vspace{0.5cm}

In this paper, we shall consider integrability condition on the
equations of motions and the action function as well in order to
obtain the path integral quantization of singular Lagrangian with
linear velocities.

\section{The Hamilton-Jacobi method}

In this section, we shall briefly review the Hamiltonian
formulation of constrained systems [3-6]. The starting point of
this method is to consider the Lagrangian $L =L (q_{i}, {\dot
q_{i}}, t), \;i= 1, ..., n$, with the  Hessian matrix
\begin{equation}
A_{ij} = \frac{\p ^{2} L}{\p {\dot q_{i}}\p {\dot
q_{j}}},\;\;\;\;i,\;j=1,...,n,
\end{equation}
of rank $(n-r)$, $r< n$. Then $r$ momenta are dependent. The
generalized momenta $p_{i}$ corresponding to the generalized
coordinates $q_{i}$ are defined as
\begin{eqnarray}
&& p_{a}= \frac{\p L}{\p \dot q_{a}},\;\;a=1, 2, ..., n-r,\\
&& p_{\m}= \frac{\p L}{\p \dot x_{\m}},\;\;\m = n-r + 1,..., n,
\end{eqnarray}
where $q_{i}$ are divided into two sets, $q_{a}$ and $x_{\m}$.
Since the rank of the Hessian matrix is $(n- r)$, one solve Eq.
(10) for ${\dot q_{a}}$ as
\begin{equation}
{\dot q_{a}} = {\dot q_{a}}(q_{i}, {\dot x_{\m}}, p_{a}; t).
\end{equation}
Substituting Eq. (12) into Eq. (11), we get
\begin{equation}
p_{\m}= -H_{\m}(q_{i}, {\dot x_{\m}}, p_{a}; t).
\end{equation}
The canonical Hamiltonian $H_{0}$ reads
\begin{equation}
H_{0}= p_{a} {\dot q_{a}} + p_{\m} \dot{x_{\m}}|_{p_{\n}=-H_{\n}}-
L(t, q_i, \dot{x_{\n}}, \dot{q_{a}}),\;\;\;\m,\;\n= n-r+1,...,n.
\end{equation}
The set of Hamilton-Jacobi partial differential equations [HJPDE]
is expressed as [3-6]
\begin{equation}
H^{'}_{\a}\left(x_{\bt}, q_a, \frac{\p S}{\p q_a},\frac{\p S}{\p
x_{\a}}\right) =0,\;\;\;\a,\;\bt=0,n-r+1,..., n,
\end{equation} where\bn
&&{H'}_{0}= p_{0} +
H_{0},\\
&& H^{'}_{\m}= p_{\mu}+  H_{\m}, \en we define $p_{\bt}= {\p
S}[q_{a};x_{\a}]/{\p x_{\bt}}$ and $p_{a}= {\p
S}[q_{a};x_{\a}]/{\p q_{a}}$ with $x_{0} = t$ and $S$ being the
action. The equations of motion are obtained as total differential
equations in many variables as follows [3-6]: \bn &&dq_a=\frac{\p
H^{'}_{\a}}{\p p_a}dx_{\a},\;\;\;\;\;\;\;\;\;
dp_a= -\frac{\p H^{'}_{\a}}{\p q_a}dx_{\a},\\
&&dp_{\bt}=-\frac{\p H^{'}_{\a}}{\p x_{\bt}}dx_{\a},\\
&&\label{eq40}dz=\left(-H_{\a}+ p_a \frac{\p H^{'}_{\a}}{\p
p_a}\right)dx_{\a},\en where $z=S(x_{\a};q_a)$. The analysis of a
constrained system is reduced to solve equations (10) with
constraints $H^{'}_{\a}=0$. Variation of constraints (7)
considering equations (10) may vanish identically or give rise to
new constraints. In the case of new constraints we should consider
their variations also. Repeating this procedure, one may obtain a
set of constraints such that all the variations vanish.
Simultaneous solutions of canonical equations with all these
constraints provide the solution of a singular system. In fact, in
references [7], the integrabilty conditions for equations (10, 11)
are discussed without considering the integrabilty conditions of
the action function.

\section{Completely and Partially Integrable Systems}

As was clarified, that the equations (10-12) are obtained as total
differential equations in many variables, which require the
investigation of integrabilty conditions. To achieve this goal we
define the linear operators $X_{\a}$ which corresponds to total
differential equations (6,7) as \bn X_{\a} f(t_{\bt}, q_{a},
p_{a}, z) &&= \frac{\p f}{\p t_{\a}} + \frac{\p H^{'}_{\a}}{\p
p_a}\frac{\p f}{\p q_a}- \frac{\p H^{'}_{\a}}{\p q_a}\frac{\p
f}{\p p_a} \nonumber\\&&+ (-H_{\a}+ p_a \frac{\p
H^{'}_{\a}}{\p p_a})\frac{\p f}{\p z},\nonumber\\
&&= [H^{'}_{\a}, f] - \frac{\p f}{\p z} H^{'}_{\a},\\
&&\a, \bt=0,n-r+1,...,n, a=1,...,n-r,\nonumber \en where the
commutator $[ , ]$ is the square bracket (for details, see the
appendix).

$\it{\bf lemma}$. A system of total differential equations (10-12)
is integrable if and only if
\begin{equation}
\{H^{'}_{\a}, H^{'}_{\bt}\} =0,\;\;\; \forall\; \a, \;\bt,
\end{equation}
where the commutator $\{ , \}$ is the Poisson bracket ( for
details see references [8, 11, 12]).

Equations (14) are the necessary and sufficient conditions that
the system (10-12) of total differential equations be completely
integrable and we call this system as ${\it Completely~
Integrable~ Model}$ . However, equations (10,11) form here by
themselves a completely integrable system of total differential
equations. If these are integrated, then only simple quadrature
has to be carried out in order to obtain the action [7].

On the other hand, we must emphasis that the total differential
equations (10,11) can be very well be completely integrable
without (14) holding and therefore without the total system
(10-12) being integrable and we call this system as ${\it
Partially~ Integrable~ Model}$. In fact, if $\{ H^{'}_{\bt},
H^{'}_{\a}\} = F_{m}(t, t_{\mu})$, where $F_{m}$ are functions of
$t_{\a}$ and $m$ is integer, then the total differential equations
(10, 11), will be integrable [8, 11, 12].

If the set of equations (10-12) is integrable, then one can obtain
the canonical action function (12) in terms of the canonical
coordinates. In this case, the path integral representation may be
written as [10] \bn
&&\Psi({q'}_a,{t'}_{\a};q_a,t_{\a})=\int_{q_a}^{{q'}_a}~Dq^{a}~Dp^{a}\times
\nonumber\\&&\exp i \left\{\int_{t_{\a}}^{{t'}_{\a}}\left[-H_{\a}+
p_a\frac{\p H^{'}_{\a}}{\p
p_a}\right]dt_{\a}\right\},\nonumber\\&&a=1,...,n-r,\;\; \a=0,
n-r+1,...,n. \en

\section{The Model}

In this section we would like to investigate singular with linear
velocities by using the Hamilton-Jacobi method [3-6, 8, 12], in
order to obtain the path integral quantization for these systems.
Let us consider the following linear Lagrangian [9, 13]
\begin{equation}
\label{eq42}L =a_i(q_j)\dot {q_i} - V(q_j),
\end{equation}
where $a_i(q_j)$ and $V(q_j)$ are continues functions of $q_j$.

The generalized canonical momentum corresponding to this
Lagrangian are given by
\begin{equation}
\label{eq43} p_{i}= \frac{\p L}{\p \dot {q_i}}=a_i(q)=-H_i.
\end{equation}

The primary constraints are
\begin{equation}
\label{eq44} H_{i}^{'}=p_{i}-a_i.
\end{equation}
The canonical Hamiltonian $H_0$ is given by:
\begin{equation}
\label{eq45} H_{0}=p_{i}\dot {q_i}-L=V(q_j).
\end{equation}
The corresponding HJPDEs are
\begin{eqnarray}
&& \label{eq46} H_{0}^{'}=p_{0}+H_0=p_0+V(q_j)=0,\\
&& \label{eq47} H_{i}^{'}=p_{i}+H_i=p_i-a_i(q_j)=0.
\end{eqnarray}
The equations of motion are obtained as total differential
equations follows:
\begin{eqnarray}
&& \label{eq48} dq_{i}=\frac{\partial H_{0}^{'}}{\partial
p_{i}}dt+
\frac{\partial H_{j}^{'}}{\partial p_{i}}dq_j=dq_i,\\
&& \label{eq49} dp_{i}=-\frac{\partial H_{0}^{'}}{\partial
q_{i}}dt-\frac{\partial H_{j}^{'}}{\partial
q_{i}}dq_j=-\frac{\partial V(q)}{\partial q_{i}}dt+\frac{\partial
a_{j}(q)}{\partial q_{i}}dq_j.
\end{eqnarray}
To check wether the set of equations (\ref{eq48}) and (\ref{eq49})
are integrable or not, let us consider the total variation of
(\ref{eq47}). In fact
\begin{eqnarray}
&& dH_{i}^{'}=dp_{i}-da_i(q)=0\nonumber\\
&& \label{eq51} ~~~~~=-\frac{\partial V(q)}{\partial
q_{i}}dt+\frac{\partial a_{j}(q)}{\partial q_{i}}dq_j-da_i(q)
\end{eqnarray}
So, we have
\begin{equation}
\label{eq52} \frac{\partial a_i(q)}{\partial
q_{j}(q)}dq_j-\frac{\partial a_{j}(q)}{\partial
q_{i}}dq_j=-\frac{\partial V(q)}{\partial q_{i}}dt,
\end{equation}
or
\begin{equation}
\label{eq53} \dot{q_j}=-f_{ij}^{-1}~\frac{\partial V(q)}{\partial
q_{i}},
\end{equation}
where the anti-symmetric matrix $f_{ij}$ is given by
\begin{equation}
\label{eq54} f_{ij}=\frac{\partial a_i(q)}{\partial
q_{j}(q)}-\frac{\partial a_{j}(q)}{\partial q_{i}}.
\end{equation}
Making use of (\ref{eq40}), (\ref{eq46}) and (\ref{eq47}), we can
write the canonical action integral as
\begin{equation}
\label{eq55} S=\int{a_i}dq_i-\int{V(q_j)}dt.
\end{equation}
In fact,
\begin{equation}
\label{eq56} \int{d(a_iq_i)}=a_iq_i=\int{a_i}dq_i+\int{q_ida_i}.
\end{equation}
So
\begin{equation}
\label{eq57}
S=\frac{1}{2}a_iq_i-\frac{1}{2}\int{\left[q_jd{a_i}-a_idq_i+2V(q)dt\right]}.
\end{equation}
Making use of (\ref{eq51}) and (29), the action function becomes
\begin{equation}\label{eq58}
S=\frac{1}{2}a_iq_i-\frac{1}{2}\int{\left[\left(2V-q_j\frac{\partial
V}{\partial q_i}\right)dt+\left(\frac{\partial a_j}{\partial
q_i}q_j-a_i\right)dq_i \right]}.
\end{equation}
Now, assuming that the functions $a_i(q_j)$ and $V(q_j)$ satisfy
the following conditions
\begin{eqnarray}
&& 2V=\frac{\partial V}{\partial q_i}q_j,~~~~~~a_i=\frac{\partial
a_j}{\partial q_i}q_j\nonumber,
\end{eqnarray}
we obtain the integrable action as follows:
\begin{equation}\label{eq59}
S=\frac{1}{2}a_iq_i+c,
\end{equation}
where $c$ is some constant. \vspace{0.2cm}
\newline
To obtain the path integral quantization for the singular
Lagrangian (16), we have three different cases,

${\it Case ~1}$: If the inverse of the matrix $f_{ij}$ exists,
then we can solve all the dynamics $q_{i}$. In this case the path
integral $\Psi$ is given by
\begin{equation}\label{eq60}
\Psi=\int{\prod_{i=1}^{n}}dq_ie^{i\left(\frac{1}{2}a_iq_i\right)}.
\end{equation}
${\it Case~ 2}$: If the rank of the matrix $f_{ij}$ is $n-R$, then
we can solve the dynamics $q_{l}$ in terms of independent
parameters $(t, q_{\a})~, \a=1, 2, ...,R$. In this case the path
integral $\Psi$ is calculated as
\begin{equation}\label{eq61}
\Psi=\int{\prod_{l=1}^{n-R}}dq_{l}e^{i\left(\frac{1}{2}a_iq_i\right)},~~~~~i=1,
...,n.
\end{equation}

${\it Case ~3}$: If $ a_{i}(q_j)$ are constants, then the path
integral $\Psi$ is given by
\begin{equation}
\Psi = e^{i \int_{q'}^{q''} (a_i~ dq_i)} = e^{i a_i({q_i}^{''}-
{q_i}^{'})},
\end{equation}
which satsifies the "Wheeler-DeWitt" equation
\begin{equation}
(p_i -a_i)\Psi=0.
\end{equation}
This result coincide  with the results obtained in reference [17]
by using the naively applied Batalin, Fradkin, Vilkovisky
procedure [18].

\section{Examples}
As a first example, we consider the following linear (singular)
Lagrangian [9]
\begin{equation}
\label{eq62}L=q_1\dot{q_1}+q_2\dot{q_2}+q_3\dot{q_1}-q_1\dot{q_3}-V,
\end{equation}
where the potential $V$ is given by
\begin{equation}
\label{eq63}V=2q_1q_3-\frac{1}{2}q_{3}^{2}.
\end{equation}
The functions $a_i (i=1, 2, 3)$ are
\begin{equation}
\label{eq64}a_1=q_1+q_3,~~~~~~~a_2=q_{2},~~~~~~~a_3=-q_1.
\end{equation}
Using (\ref{eq43}), the generalized momenta corresponding to this
Lagrangian are:
\begin{eqnarray}
&&p_1=a_1=q_1+q_3=-H_1\nonumber,\\
&&p_2=a_2=q_{2}=-H_2\nonumber,\\
&&p_3=a_3=-q_1=-H_3\nonumber.
\end{eqnarray}
By (\ref{eq44}) the primary constraints are given as
\begin{equation}
\label{eq65}H_{1}^{'}=p_1-q_1-q_3,~~~~H_{2}^{'}=p_2-q_{2},~~~~H_{3}^{'}=p_3+q_1.
\end{equation}
Equation (\ref{eq45}) gives the canonical Hamiltonian $H_0$ as
\begin{equation}
\label{eq66}H_{0}=V(q)=2q_1q_3-\frac{1}{2}q_{3}^{2}.
\end{equation}
Now using (\ref{eq49}), the equations of motion read as
\begin{eqnarray}
&&\label{eq67}dp_1=-2q_3dt+dq_1-dq_3,\\
&&\label{eq68}dp_2=dq_{2},\\
&&\label{eq69}dp_3=\left(-2q_1+q_3\right)dt+dq_1.
\end{eqnarray}
The matrix $f_{ij}$ defined in (\ref{eq54}) is given
by
\begin{equation}
\label{eq70}f_{ij}=\left(%
\begin{array}{ccc}
  0 & 0 & 2 \\
  0 & 0 & 0 \\
  -2 & 0 & 0 \\
\end{array}%
\right).
\end{equation}
Making use of  (\ref{eq51}), we can obtained the equation of
motion for $q_1, q_2$ and $q_3$ respectively as
\begin{equation}
\label{eq71}dq_3+q_3dt=0,~~~~~ 0=0,~~~~~
-2dq_1+\left(2q_1-q_3\right)dt=0.
\end{equation}
The integrable action function is calculated as
\begin{equation}
\label{eq72}S=\frac{1}{2}\left(q_{1}^{2}+q_{2}^{2}\right)+c.
\end{equation}
Making use of equation (34) and equation (47), the path integral
for the model is given by
\begin{equation}
\label{eq73}\Psi =\int
dq_{1}dq_{3}e^{i\frac{1}{2}\left(q_{1}^{2}+q_{2}^{2}\right)}.
\end{equation}

\vspace{0.3cm}

As a second example, let us consider the following linear
(singular) Lagrangian [9]
\begin{equation}
\label{eq74}L=\left(q_2+q_3\right)\dot{q_1}+q_4\dot{q_3}-V(q),
\end{equation}
where the potential $V(q)$ is given by
\begin{equation}
\label{eq75}V(q)=-\frac{1}{2}\left(q_{4}^{2}-2q_2q_3-q_{3}^{2}\right).
\end{equation}
The functions $a_i (i=1, 2, 3, 4)$ are
\begin{equation}
\label{eq76}a_1=q_2+q_3,~~~~~a_2=0,~~~~~a_3=q_4,~~~~~~a_4=0.
\end{equation}
Using (\ref{eq43}), the generalized momenta corresponding to this
Lagrangian are:
\begin{eqnarray}
&&p_1=a_1(q)=q_2+q_3=-H_1\nonumber,\\
&&p_2=a_2=0=-H_2\nonumber,\\
&&p_3=a_3=q_4=-H_3\nonumber,\\
&&p_4=a_4=0=-H_4\nonumber.
\end{eqnarray}
The primary constraints are given as
\begin{equation}
\label{eq77}H_{1}^{'}=p_1-q_2-q_3,~~~~H_{2}^{'}=p_2,~~~~H_{3}^{'}=p_3+q_4,~~~~H_{4}^{'}=p_4.
\end{equation}
Equation (\ref{eq45}) gives the canonical Hamiltonian $H_0$ as
\begin{equation}
\label{eq78}H_{0}=V(q)=-\frac{1}{2}\left(q_{4}^{2}-2q_2q_3-q_{3}^{2}\right).
\end{equation}
Now making use of (\ref{eq49}), the equations of motion read as
\begin{eqnarray}
&&\label{eq79}dp_1=0,\\
&&\label{eq80}dp_2=-q_3dt + dq_{1},\\
&&\label{eq81}dp_3=-\left(q_2+q_3\right)dt+dq_1,\\
&&\label{eq82}dp_4=q_4dt+dq_3.
\end{eqnarray}
The matrix $f_{ij}$ defined in (\ref{eq54}) is given by
\begin{equation}
\label{eq83}f_{ij}=\left(%
\begin{array}{cccc}
  0 & 1 & 1 & 0 \\
  -1 & 0 & 0 & 0 \\
  -1 & 0 & 0 & 1 \\
  0 & 0 & -1 & 0 \\
\end{array}%
\right)
\end{equation}
The $S$-action function is calculated as
\begin{equation}
\label{eq85}S=\frac{1}{2}\left[(q_{2}+q_{3})q_1+q_4q_3\right]+c.
\end{equation}
Making use of (33) and (59), the path integral for the model is
calculated as
\begin{equation}
\label{eq86} \Psi =\int
dq_{1}dq_{2}dq_{3}dq_{4}e^{i\frac{1}{2}\left[(q_{2}+q_{3})q_1+q_4q_3\right]}.
\end{equation}

An important point to be specified here is that, in the Jackiw's
method treatment of the above model [19], the path integral is
obtained by introducing $\delta$ function in the measure of the
integral, while in the canonical path integral method [10], the
path integral quantization is obtained directly as an integration
over the canonical variables $q_{1}, ~q_{2}, ~q_{3}, ~q_{4}$
without any need to use these $\delta$ functions.

\section{Conclusion}
In this work we have investigated constrained systems using the
Hamilton-Jacobi method for Lagrangians with linear velocities. The
equations of motion are obtained from the integrabilty conditions
and the number of independent parameters $(multi-times)$ are
determined from the rank of matrix $f_{ij}$. Besides the
integrable action is obtained from the integrability conditions,
which leads us to obtain the path integral quantization for the
singular Lagrangians with linear velocities directely as an
integration over the independent dynamical variables without any
need to use $\delta$ functions as given in the Faddeev, Jackiw
method [19].
\section{Square brackets and Poisson brackets}
In this appendix we shall give a brief review on two kinds of
commutators: the square and the Poisson brackets.

The square bracket is defined as
\begin{equation}
[F, G]_{q_{i}, p_{i}, z}= \frac{\p F}{\p p_{i}} \frac{\p G}{\p
q_{i}}- \frac{\p G}{\p p_{i}}\frac{\p F}{\p q_{i}} + \frac{\p
F}{\p p_{i}}(p_{i} \frac{\p G}{\p z}) - \frac{\p G}{\p
p_{i}}(p_{i} \frac{\p F}{\p z}).
\end{equation}
The Poisson bracket is defined as
\begin{equation}
{\{f, g\}}_{q_{i}, p_{i}}= \frac{\p f}{\p p_{i}} \frac{\p g}{\p
q_{i}}- \frac{\p g}{\p p_{i}}\frac{\p f}{\p q_{i}}.
\end{equation}
According to above definitions, the following relation holds
\begin{equation}
[H^{'}_{\a}, H^{'}_{\bt}]= \{H^{'}_{\a}, H^{'}_{\bt}\}.
\end{equation}

\end{document}